\documentclass[aps,pre,twocolumn,superscriptaddress,showpacs,showkeys,longbibliography]{revtex4-1}
\usepackage[english]{babel}

\usepackage{graphicx}
\usepackage{grffile}
\usepackage{chngcntr}
\usepackage{dcolumn}
\usepackage{bm}
\usepackage{float}
\usepackage{mathtools}
\usepackage{amsmath}

\usepackage{amssymb}

\usepackage{etoolbox}
\usepackage{tikz}

\newrobustcmd*{\mytriangle}[1]{\tikz{\filldraw[draw=#1,fill=#1] (0,0) --
(0.2cm,0) -- (0.1cm,0.2cm);}}

\begin{document}

\preprint{APS/123-QED}

\title{Community detection in networks using graph embeddings}






\date{\today}
\author{Aditya Tandon}
\affiliation{Luddy School of Informatics, Computing and Engineering, Indiana University}

\author{Aiiad Albeshri}
\affiliation{Department of Computer Science, Faculty of Computing and Information Technology
King Abdulaziz University, Jeddah 21589, Kingdom of Saudi Arabia}

\author{Vijey Thayananthan}
\affiliation{Department of Computer Science, Faculty of Computing and Information Technology
King Abdulaziz University, Jeddah 21589, Kingdom of Saudi Arabia}

\author{Wadee Alhalabi}
\affiliation{Department of Computer Science, Faculty of Computing and Information Technology
King Abdulaziz University, Jeddah 21589, Kingdom of Saudi Arabia}

\author{Filippo Radicchi}
\affiliation{Indiana University Network Science Institute (IUNI)}
\affiliation{Luddy School of Informatics, Computing and Engineering, Indiana University}

\author{Santo Fortunato}
\affiliation{Indiana University Network Science Institute (IUNI)}
\affiliation{Luddy School of Informatics, Computing and Engineering, Indiana University}

\begin{abstract}
Graph embedding methods are becoming increasingly popular in the machine learning community, where they are 
widely used for tasks such as node classification and link prediction. Embedding graphs in geometric spaces 
should aid the identification of
network communities as well, because nodes in the same community should be projected close to each other in the geometric space, where they can be detected via standard data clustering algorithms. In this paper, we test the ability of several graph embedding techniques to detect communities on benchmark graphs. We compare their performance against that of traditional community detection algorithms. We find that the performance is comparable, if the parameters of the embedding techniques are suitably chosen. However, the optimal parameter set varies with the specific features of the benchmark graphs, like their size, whereas popular community detection algorithms do not require any parameter. So it is not possible to indicate beforehand good parameter sets for the analysis of real networks. This 
finding, along with the high computational cost of embedding a network and grouping the points, suggests
that, for community detection, current embedding techniques do not represent an improvement over network clustering algorithms. 

\end{abstract}


\maketitle






\section{Introduction}

Community structure is a common feature of many complex systems that can be represented as networks. A community, or cluster, is traditionally conceived as a set of nodes having a substantially higher probability to be linked to each other than to the rest of the graph. Detecting communities is a classic task in network analysis, that helps to uncover important structural and functional information of networks~\cite{fortunato10, fortunato16, porter09}. 
It is an unsupervised classification problem, and as such it is ill-defined. Nevertheless a huge number of algorithms have been developed over the past two decades. 

Most techniques rely on a characterization of the structural properties of communities, typically via the density of internal links. Another approach is \textit{spectral clustering}~\cite{luxburg06}, where the nodes of the network are projected in a $k$-dimensional Euclidean space by using the top (or bottom) $k$ eigenvectors of graph matrices, like the adjacency matrix or, more frequently, the Laplacian. This way the network is \textit{embedded} in a geometric space, and, if it has a pronounced community structure, clusters of nodes appear as groups of points which are close to each other, and well separated from the points in the other groups. Such concentrations of points can be identified via standard data clustering techniques.

In the last decade several approaches have been developed to embed graphs in high-dimensional geometric spaces, while preserving some of their properties~\cite{Goyal2018}. Such embedding techniques have proven to be useful to solve important tasks with many potential applications, e.g., node classification, link prediction, graph visualization~\cite{pereda19}. Node classification aims at determining the label of nodes based on other labeled nodes and the topology of the
network. Link prediction aims at predicting missing
links (e.g., because of incomplete data) or links that are likely to occur in the future.

However, it is unclear whether such embedding strategies make the identification of communities easier than by applying traditional network clustering algorithms. Here we address this specific issue.
We adopt
a broad collection of embedding techniques to project in vector spaces artificial networks with planted communities. We 
use
the \textit{Lancichinetti-Fortunato-Radicchi (LFR) benchmark graphs}~\cite{lancichinetti08}, 
which are regularly employed to test clustering algorithms. The resulting distributions of geometric points are divided into clusters by using the k-means algorithm~\cite{macqueen67}.

Our analysis reveals that default values of the parameters for the embedding methods, recommended 
based on their optimal performance in other tasks,
lead to comparable performance as traditional non-parametric community detection methods, though performance tends to degrade for larger networks. A careful exploration of the parameter space often allows superior performance than standard clustering algorithms, but the results are very sensitive to little variations in the parameters' values
and we could not find specific sets of values leading to good solutions in most cases.

\section{Methods}
\subsection{LFR benchmark}
\label{sec:LFR}

The LFR benchmark is characterized by power-law distributions of degree and community size, reflecting the heterogeneity of these two variables in real graphs~\cite{lancichinetti08}. 
The parameters needed to generate the graphs are the number of nodes $N$, the exponents of the distributions of degree ($\tau_1$) and community size ($\tau_2$), the average degree $k$, the maximum degree $k_{max}$, the extremes of the range of community sizes $c_{min}$ (lower) and $c_{max}$ (upper), and the mixing parameter $\mu$. The latter indicates how pronounced the community structure is. If $\mu=0$ clusters are disjoint from each other, i.e., all links fall within communities, and are easily detectable.  For $\mu=1$ links fall solely between clusters, which are not communities in the traditional sense of cohesive subsets of nodes, albeit they might be still detectable with particular techniques~\cite{peixoto20}. Normally the parameters are fixed at the onset, except $\mu$ which is varied to explore different strength of communities.
For our tests we used two different sets of parameters, mostly to explore how the clustering performance of embedding techniques is affected by the network size. The two sets are:

\begin{itemize}
    \item $N = 1,000$,
    $\tau_1 = 2$,
    $\tau_2 = 3$,
    $\bar{k} = 20$,
    $k_{max} = 50$,
    $c_{min} = 10$,
    $c_{max} = 100$.
    \item $N = 10,000$,
    $\tau_1 = 2$,
    $\tau_2 = 3$,
    $\bar{k} = 20$,
    $k_{max} = 200$,
    $c_{min} = 10$,
    $c_{max} = 1,000$.
\end{itemize}

The similarity between the planted partition of the benchmark and the one found by the clustering algorithm can be measured in various ways. We opted for the \textit{normalized mutual information} (NMI)~\cite{ana03}, a measure borrowed from information theory, which is regularly used in this type of tests.
For each value of $\mu$ we generated $20$ configurations of the benchmark and report the average value of the NMI for such set of configurations as a function of $\mu$.

For a thorough assessment of clustering performance, we also carried out tests on the popular benchmark by Girvan and Newman~\cite{girvan02}, where communities have the same size and nodes have the same degree. The results are reported in Appendix A and confirm the conclusions of our work.

\subsection{Data clustering}

After obtaining the embedding, $k$-means clustering is adopted to group the points into clusters.  $k$-means is very popular in data clustering. To make sure our message is not strongly depending on the choice of the specific clustering method we used Gaussian Mixture Models~\cite{reynolds09} as well, which leads to similar results.
$k$-means clustering minimizes the squared distance between each data point (corresponding to a network node) and its centroid, which is a virtual point representing its cluster. 

In the high-dimensional spaces where the network is embedded, the distance between points becomes less and less useful and the concept of proximity may not be meaningful, as different notions of distances may select different neighbors for the same point.
For instance, the ratio of the distances of the nearest and farthest neighbors
to a given target in a high dimensional space is almost one for a wide variety of data
distributions and distance functions~\cite{beyer99}. For this reason, in our calculations we have used both the standard $k$-means, based on the \textit{Euclidean distance}, as well as spherical $k$-means~\cite{Dhillon2001}, which uses the \textit{spherical distance}. Given two points $i$ and $j$ identified by the vectors of coordinates ${\bf x}_i=(x_i^1,x_i^2,x_i^3,\dots,x_i^{d-1},x_i^d)$ and ${\bf x}_j=(x_j^1,x_j^2,x_j^3,\dots,x_j^{d-1},x_j^d)$, where $d$ is the number of dimensions, the two 
distance metrics are defined as follows:
\begin{itemize}
    \item \textit{Euclidean distance}
    \begin{equation}
        dist_E(i,j)=\sqrt{\sum_{l=1}^d (x_i^l-x_j^l)^2}.
    \end{equation}
    \item \textit{Spherical distance}
    \begin{equation}
        dist_S(i,j)=1-\frac{\sum_{l=1}^d x_i^l\,x_j^l}{\sqrt{\sum_{l=1}^d (x_i^l)^2\sum_{l=1}^d (x_j^l)^2}}.
    \end{equation}
\end{itemize}

Data clustering techniques typically require the knowledge of the number of clusters beforehand. Instead of inferring this number via some criterion, we feed the procedure with the correct number of clusters of the benchmark graphs. This way we will actually assess the optimal performance of the embedding techniques. This is a luxury that we do not have when analyzing real networks, for which the number of clusters is unknown.

To improve the quality of the clustering, for each network the procedure is run 100 times, and the partition corresponding to the minimum distance between points and their centroids is selected. 

\subsection{Community Detection}

We 
compare the cluster analysis via embedding algorithms with the performance of widely adopted methods for community detection. We specifically show the comparison with Infomap~\cite{rosvall08} and the Louvain algorithm~\cite{blondel08}, that are known to be especially accurate to identify the planted partition of LFR benchmark graphs~\cite{lancichinetti09}.
Other methods have similar performances and we report them in Appendix B.
Infomap is based on diffusion dynamics: if the graph has a pronounced community structure, a random walker will spend a lot of time within a community before finally finding a bridge taking it to another community. This way, the description of an infinitely long random walk can be reduced by using the same labels for nodes of different clusters, much like it is done in geographic maps for the names of towns belonging to different regions/states. The partition leading to the cheapest description of the random walk is the best, by construction. The method does not require any parameter. It has some shortcomings, like the inability to detect clusters smaller than a certain scale~\cite{kawamoto15} and the tendency to split non-clique structures with large diameters, such as strings
and lattices~\cite{schaub12b}. Nevertheless it is frequently used in applications. The code was taken from the \href{http://igraph.org/}{\textit{igraph}} library.

The Louvain algorithm is a fast greedy method to optimize the modularity of Newman and Girvan~\cite{newman04b}, a function that estimates the goodness of a partition of a graph in communities. It is a very popular technique but, like Infomap, it has important limitations, like the inability to detect clusters smaller than a certain scale ~\cite{fortunato07}. This is why here we consider the partition derived in the first step of the algorithm (the one with the smallest clusters), rather than the one with the largest value of modularity, which gives poor performance~\cite{fortunato16}.
We used the \href{https://github.com/taynaud/python-louvain}{\textit{python-louvain}} package for \href{networkx.org}{\textit{NetworkX}}.

\subsection{Genetic optimization}
\label{sec:GA}
Several graph embedding algorithms have multiple free parameters that have to be chosen. In the case of a single free parameter an exhaustive search can be done over the range of parameter values to pick the one giving the best clustering performance. However, for other algorithms, exhaustive search over all free parameters becomes computationally infeasible. To find the best set of parameters we use a simple genetic algorithm outlined in Ref.~\cite{Back2018} and use the python package DEAP \cite{Fortin2012} to implement the algorithm. The parameters used for the optimization procedure are:
\begin{enumerate}
    \item Number of individuals in the population = $50$.
    \item Number of generations = $20$.
    \item Mutation function: normal distribution with mean $0$ and standard deviation $3$. 
    \item Mutation probability = $0.2$.
    \item Probability of an individual being produced by crossover = $0.5$.
\end{enumerate}

\section{Results}

In this section we present the results of our tests.
Each of the following subsections reports the results for a different class of embedding techniques.

\subsection{Matrix-based Methods}

The approach consists in projecting nodes onto a high-dimensional vector space via eigenvectors of graph matrices, which is the same principle of spectral clustering~\cite{luxburg06}.

\subsubsection{Laplacian Eigenmap (LE)}

\begin{figure}
\includegraphics[width=\columnwidth]{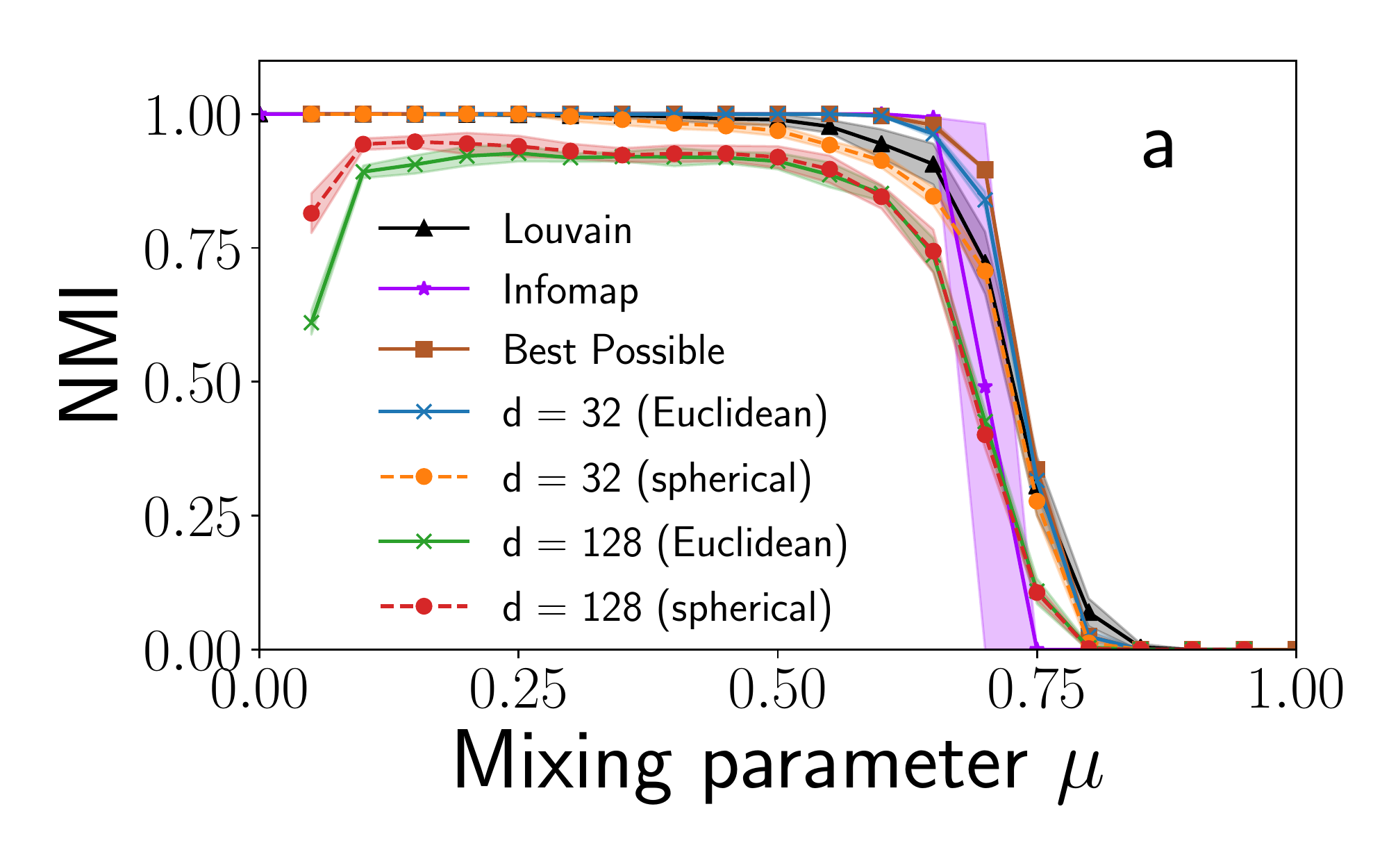}
\includegraphics[width=\columnwidth]{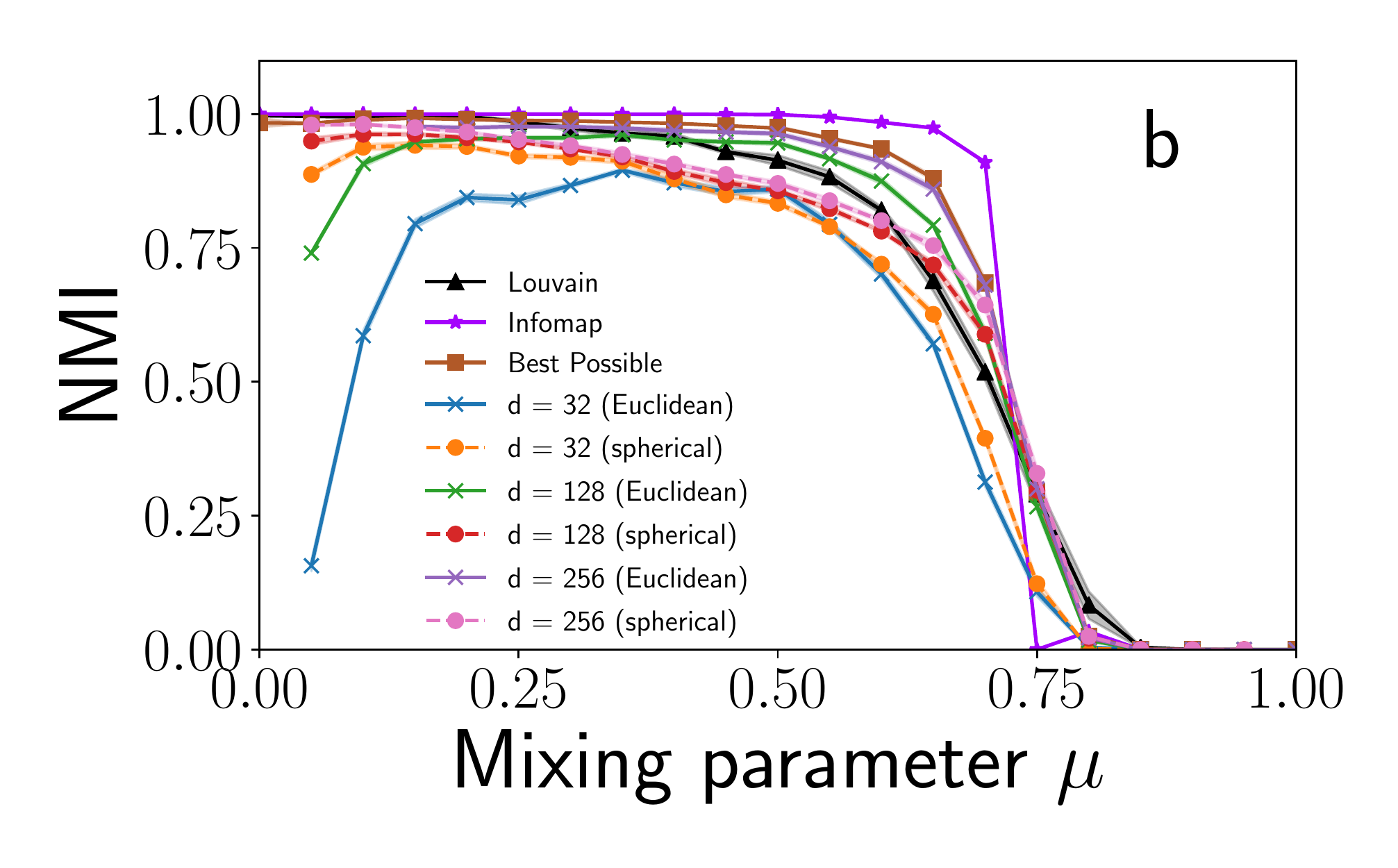}
\caption{Performance of Laplacian Eigenmap on LFR networks of 1,000 nodes (a) and 10,000 nodes (b). The NMI values are averaged over 10 realizations of the same LFR configuration for each $\mu$. The baseline performances are those of Infomap ($\star$) and Louvain (\mytriangle{black}). The shaded area of the Infomap curve in (a) reflects the large fluctuations of the NMI when the performance declines: here the partition found by the algorithm oscillates between one correlated with the planted one and the partition into one cluster, so the distribution of the NMI for those values of $\mu$ is bimodal (with peaks near 1 and 0), which gives a large variance. The spherical distance metric ($\bullet$) performs better than euclidean distance ($\textbf{x}$) in the regime of low $\mu$ and high embedding dimension $d$. The best possible performance (\rule{1.2ex}{1.2ex})
is found by doing an exhaustive search across 50 different values of $d$ ranging from 2 to 500 for each network and selecting the one with the largest NMI.}
\label{fig:LE}
\end{figure}

\textit{Laplacian Eigenmap} (LE) \cite{Belkin2003} minimizes, under some constraints, the objective function

\begin{equation}
    E_{LE} =  \sum_{ij} | \mathbf{x_i} - \mathbf{x_j}|^2 A_{ij} \; ,
\end{equation}
where $\textbf{x}_i$ is the vector indicating the position of the point representing node $i$ in the embedding and $\textbf{A}$ is the adjacency matrix of the graph, whose elements weigh the square distance between the corresponding points. Nodes that have a higher link weight will consequently be closer together in the embedding. The constraint $\textbf{x}^T \textbf{D} \textbf{x} = 1$, where ${\bf D}$ is the diagonal matrix of the degrees of the graph nodes, is added to remove an arbitrary scaling factor in the embedding. If we wish an embedding in $d$ dimensions, it can be shown that the desired one is obtained via the eigenvectors corresponding to the $d$ lowest eigenvalues (except the zero eigenvalue) of the problem ${\bf \widetilde{L}}\textbf{x} = \lambda {\bf D} \textbf{x}$, where ${\bf \widetilde{L} = D^{-\frac{1}{2}}(D-A)D^{-\frac{1}{2}}}$ is the normalized Laplacian matrix of the graph. The projection $\textbf{x}_i$ is the vector whose components are the $i$-th entries of the $d$ eigenvectors. LE has a single parameter --- the embedding dimension $d$, making it straightforward and easy to use. 
In Fig.~\ref{fig:LE} we show how LE performs for different values of $d$ (i.e., $32$, $128$, $256$) and the two $k$-means clustering procedures we have chosen. The panels correspond to the different sets of parameters of the LFR benchmark listed in Section~\ref{sec:LFR}. The performance of Infomap is generally superior, despite the fact that the latter has no parameters and ignores the number of clusters, which is derived by the algorithm itself. By doing an exhaustive optimization, for each single network (even when $\mu$ is the same) we have identified the value of $d$ giving the best performance. The resulting performance curves are a bit better than Infomap's on the smaller networks, while on the larger ones Infomap has still an edge. However, the optimal $d$-value varies with $\mu$. It is well known that, in spectral clustering, the ideal number of eigenvectors to use to obtain a good clustering typically matches the number of clusters to be found~\cite{luxburg06}. We find that this is mostly true here, but not always. Louvain has comparable perfomance as LE on LFR graphs with 1,000 nodes, while its curve worsens on the larger graphs.

We remark that for low values of $\mu$ the performance is unexpectedly poor, given that communities are well separated from each other. It turns out that in these cases some nodes of different clusters are projected close to each other, making it hard to correctly classify them (see Appendix C).

\subsubsection{Locally Linear Embedding (LLE)}

\begin{figure}
\includegraphics[width=\columnwidth]{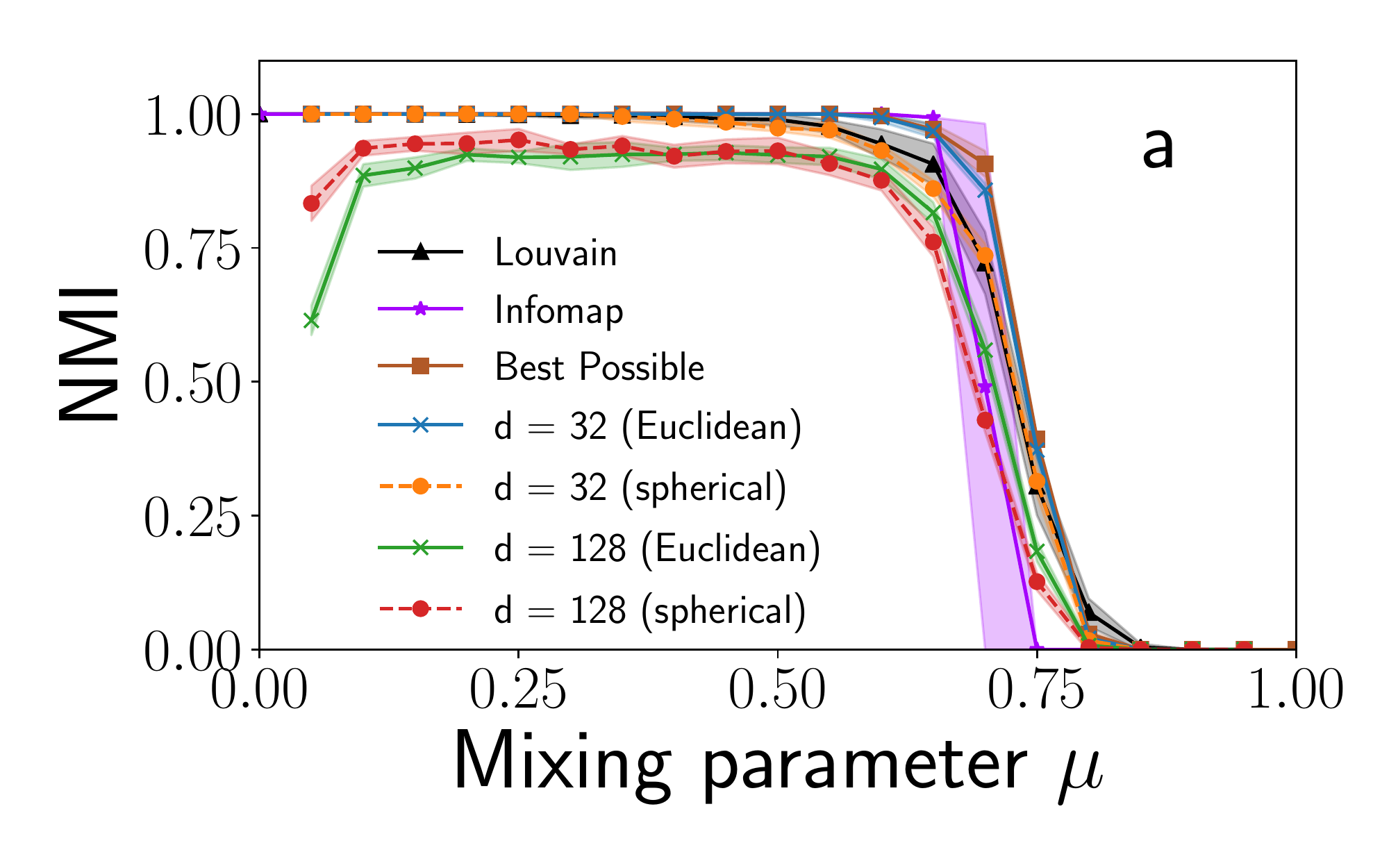}
\includegraphics[width=\columnwidth]{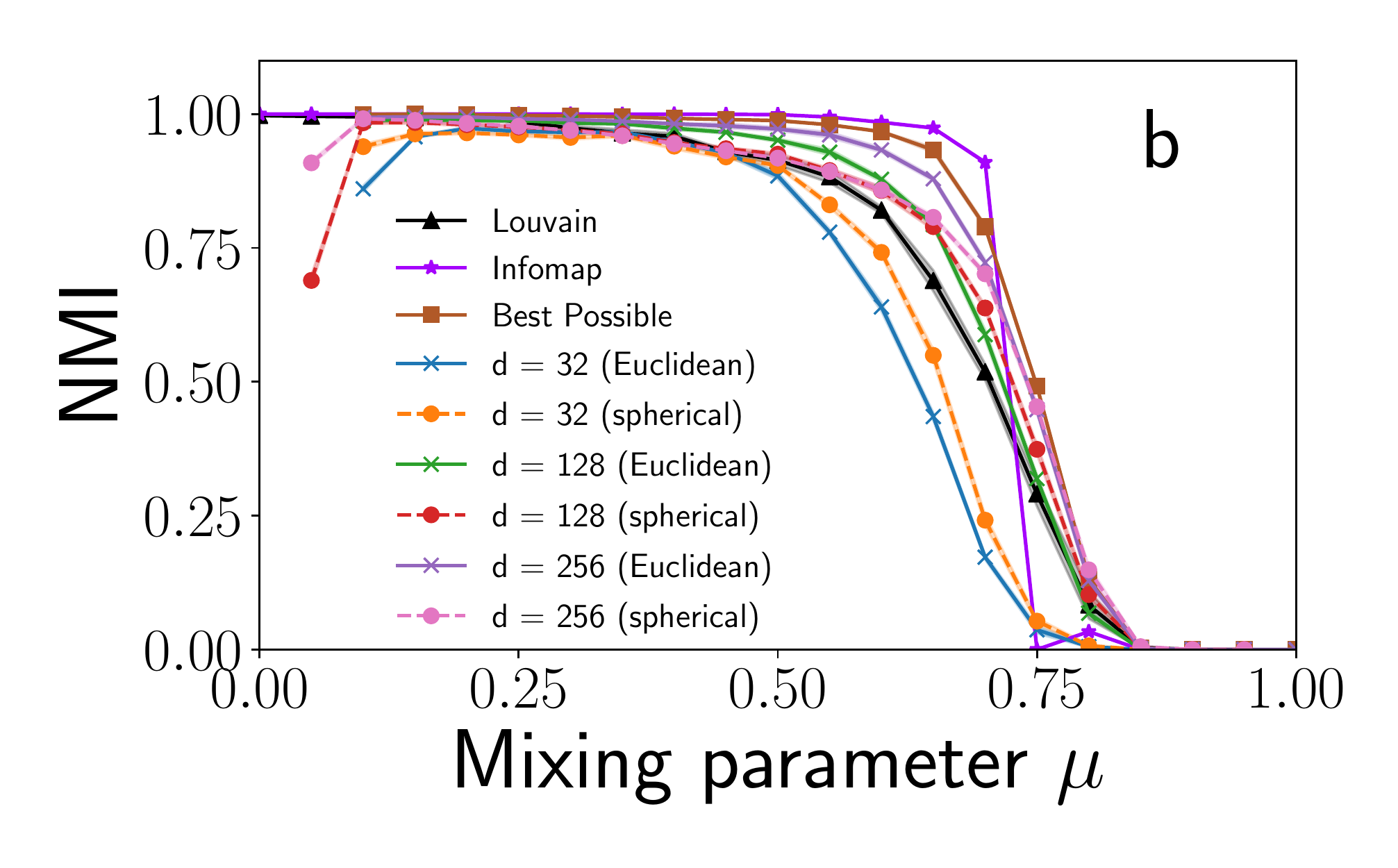}
\caption{Performance of Locally Linear Embedding on LFR networks of 1,000 nodes (a) and 10,000 nodes (b). The NMI values are averaged over 10 realizations of the same LFR configuration for each $\mu$. The baseline performances are those of Infomap ($\star$) and Louvain (\mytriangle{black}). The cosine distance metric ($\bullet$) performs better than the Euclidean distance ($\textbf{x}$) in the regime of low $\mu$ and high embedding dimension ($d$). The best possible NMI (\rule{1.2ex}{1.2ex}) is found by doing an exhaustive search across 50 different values of $d$ ranging from 2 to 500, for each network and selecting the one with the largest NMI.}
\label{fig:LLE}
\end{figure}

\textit{Locally Linear Embedding} (LLE) \cite{Rowels2000} minimizes the objective function 
\begin{equation}
    E_{LLE} =  \sum_{i}  |\mathbf{x_i} - \sum_{j} A_{ij} \mathbf{x_j}|^2 \;,
\end{equation} 
where each summand is the square distance between vectors.
Each point in the embedded space $\mathbf{x_i}$ is approximated as a linear combination of its neighbours in the original graph. To make the problem well posed, the solutions are required to be centered at the origin, i.e., $\sum_i \mathbf{x_i} = 0$, and have unit variance, i.e., $\frac{1}{N} \mathbf{x}^T \mathbf{x} = I$. With these constraints the solution is approximated by the eigenvectors corresponding to the lowest eigenvalues (disregarding the zero eigenvalue) of the matrix $\mathbf{M} = ({\bf I} - \textbf{A})^T ({\bf I} - \textbf{A})$. Like LE, LLE too has a single parameter -- the embedding dimension. 
The results are in Fig.~\ref{fig:LLE}, where we again consider up to three values for the number of dimensions (i.e., $32$, $128$, $256$) and derive the best performance curve by identifying the best $d$-value for each benchmark graph via exhaustive optimization. The conclusion is similar as for LE: LLE generally underperforms Infomap. The optimal performance curve is better but the values of the best number of dimensions vary with the network. Louvain is comparable to the best LLE curve for N=1,000, while it is worse for N=10,000.

\subsubsection{Higher-Order Preserving Embedding (HOPE)}

\textit{Higher-Order Preserving Embedding} (HOPE) \cite{Ou2016} aims to preserve the similarity between nodes. In its general formulation, the function that is optimized is $|| \mathbf{S} - \mathbf{x}\mathbf{x}^T ||$ where $\mathbf{S}$ is the similarity matrix and the matrix distance is the sum of the squares of the differences between the corresponding matrix elements. Although the authors try several different similarity metrics, we found \textit{Katz similarity}~\cite{katz53} to be the best performing one and used it in our experiments here. 
Katz similarity is defined as
\begin{equation}
    \mathbf{S}^\textrm{katz} =\beta \sum_{l = 1}^{\infty} \mathbf{A}^l 
\end{equation}
where $\beta < 1$ is a decay parameter. The other parameter $d$ is the embedding dimension. 

\begin{figure}[H]
\includegraphics[width=\columnwidth]{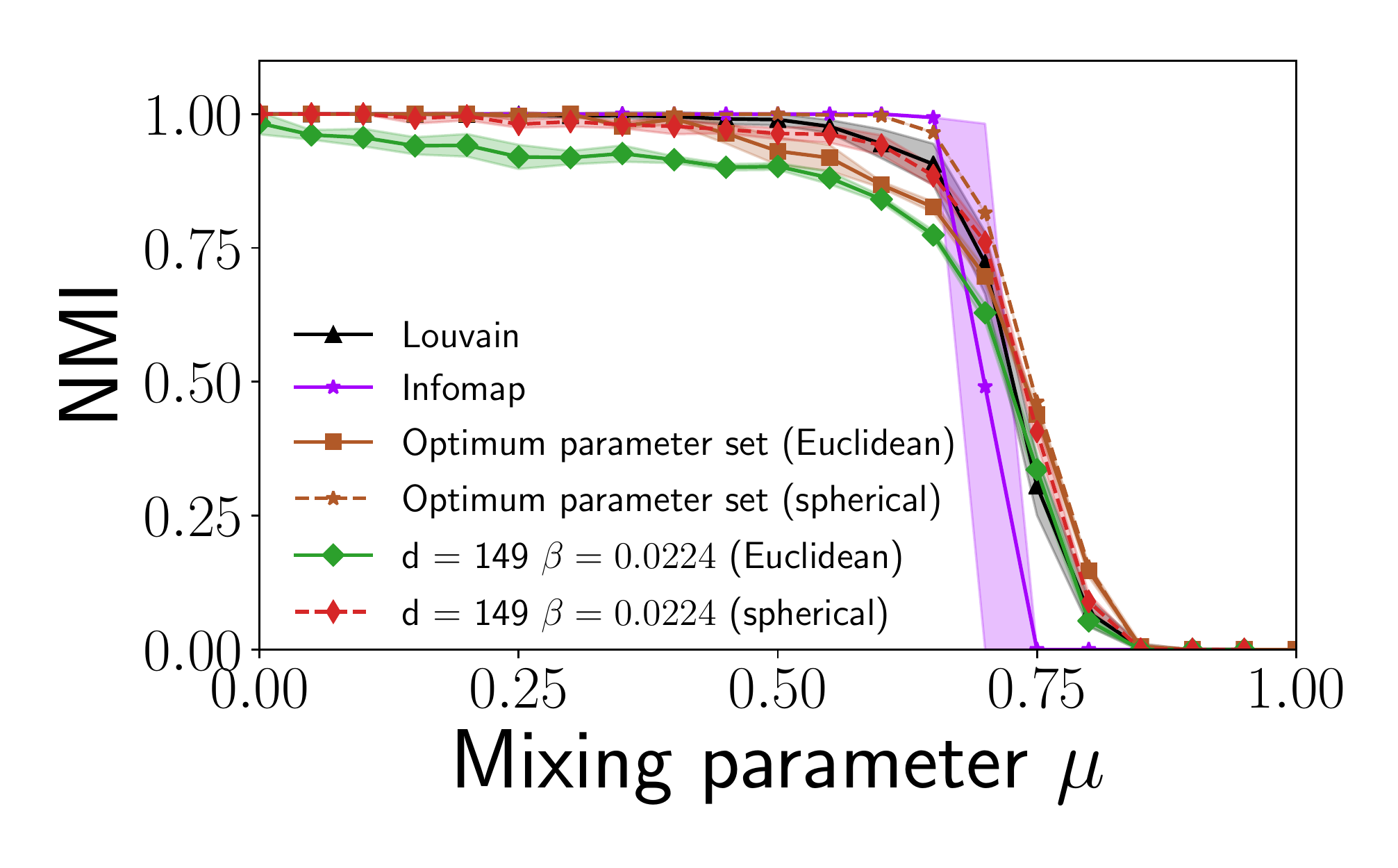}
\caption{Performance of Higher-Order Preserving Embedding on LFR networks with 1,000 nodes. The baseline performances are those of Infomap ($\star$) and Louvain (\mytriangle{black}). For each value of the mixing parameter, $\mu$, the average NMI over 10 configurations is reported. The best parameters are found by employing a genetic optimization in two different ways --- finding the best parameters for each configuration separately (brown) and finding the optimum parameter set across all values of $\mu$ (green and red, the optimal parameters are indicated) by optimizing the area under the curve. The spherical distance metric (dashed) performs better than the Euclidean metric (solid lines). }
\label{fig:HOPE}
\end{figure}

Figure~\ref{fig:HOPE} shows the results of the analysis, where we only considered the smaller LFR graphs, with $1,000$ nodes, as experiments on the larger networks have a high computational cost. The colored curves correspond to the optimal performance, which is obtained by using the genetic algorithm described in Section~\ref{sec:GA}. We followed two different approaches: 1) optimization of the parameters for each individual network; 2) optimization of the parameters for the whole curve, so that their values are fixed for every $\mu$ and graph. This is to check whether we can recommend specific pairs of values for clustering purposes. Also, we considered both $k$-means algorithms. By construction, the graph-based optimization offers a superior performance than the optimization based on the whole curve, though the difference is not big. 

The Infomap curve is better than all optimal ones until $\mu \sim 0.7$, then it undershoots them for larger $\mu$, though in that region the overall performance is quite poor because communities are well blended with each other. Louvain is comparable to both best performance curves.

\subsubsection{Modularized Nonnegative Matrix Factorization (M-NMF)}
 \textit{Modularized Nonnegative Matrix Factorization} (M-NMF) \cite{Wang2017} incorporates 
modularity~\cite{Newman2006} into the optimization function. Modularity is a function expressing how good a partition in communities is, based on the comparison between the network and randomized versions of it, which are supposed not to have community structure. The minimization of the optimization function guarantees that nodes are projected near each other if they are similar and, at the same time, if they belong to clusters of high-modularity partitions. The similarity between two nodes is the cosine similarity of vectors whose entries are the overlaps between the neighborhoods of the nodes. Although the original paper does not mention a default set of parameters, we use the following ones for the experiments in this paper - 
dimension $d = 128$, $\lambda = 0.2$, $\alpha = 0.05$, $\beta = 0.05$, $\eta = 5.0$, number of iterations $N_i = 200$.

\begin{figure}[h]
\includegraphics[width=\columnwidth]{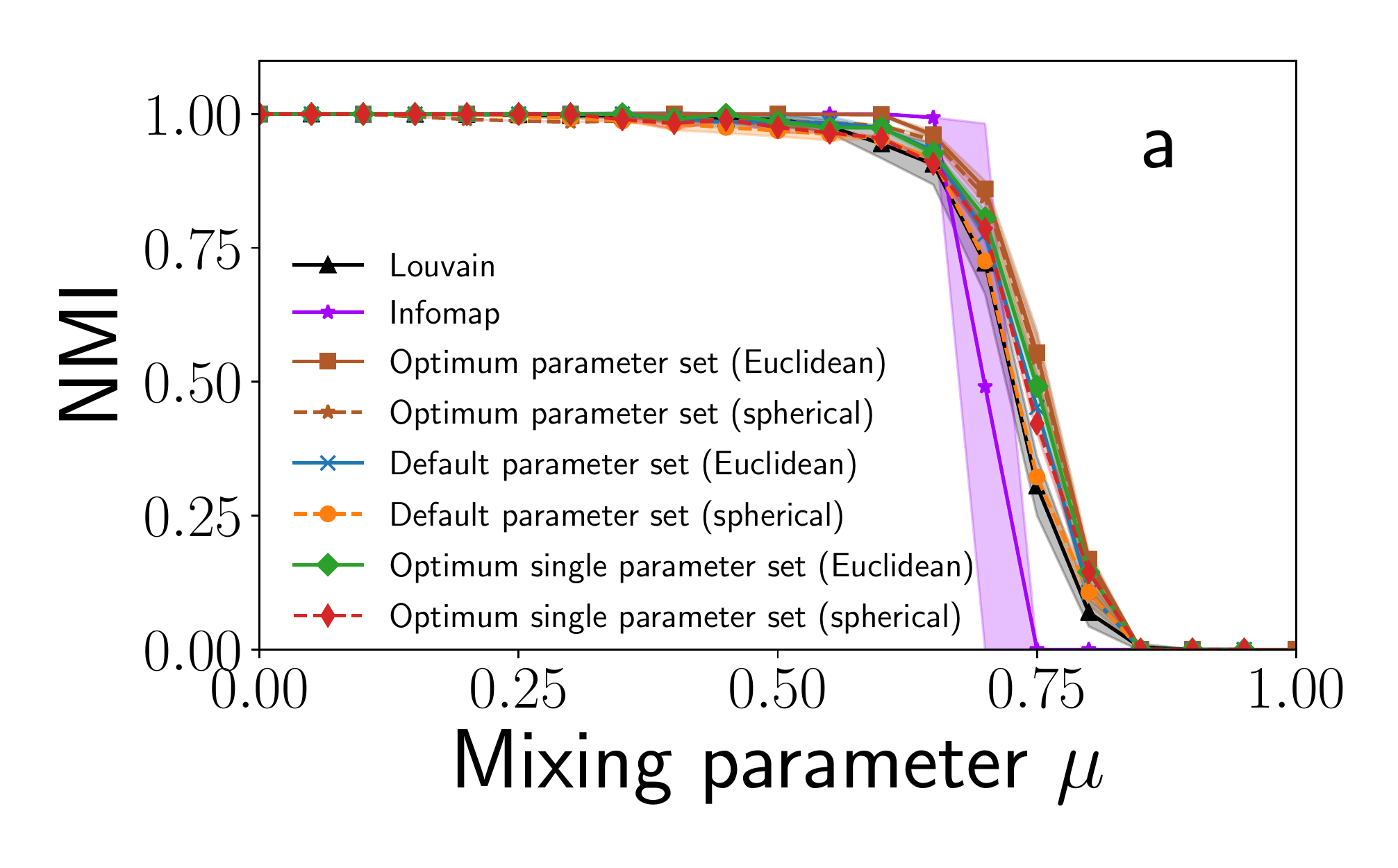}
\includegraphics[width=\columnwidth]{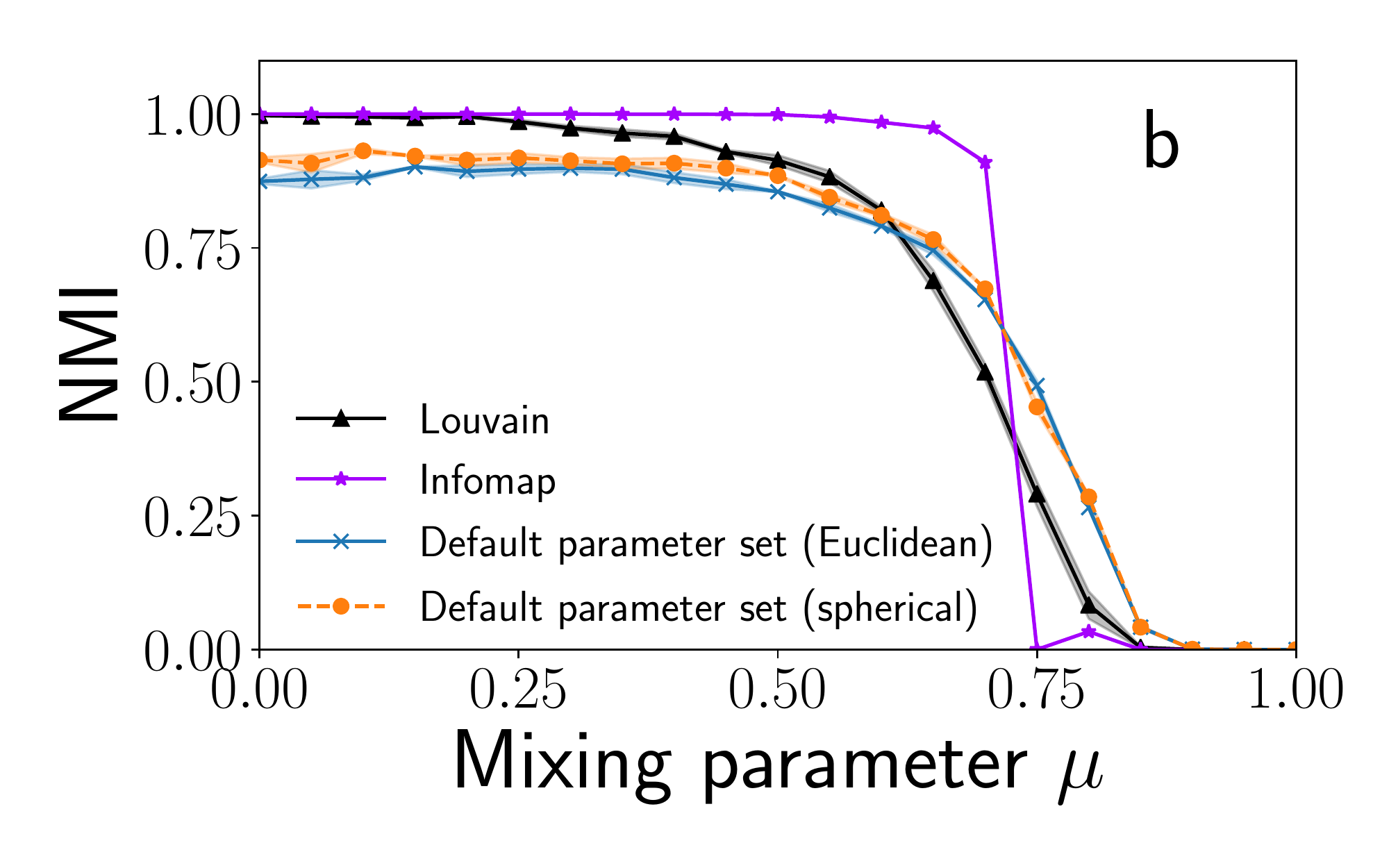}
\caption{Performance of Modularized Nonnegative Matrix Factorization for LFR benchmark graphs of 1,000 nodes (a) and 10,000 nodes (b). The baseline performances are those of Infomap ($\star$) and Louvain (\mytriangle{black}). The performance obtained by using the default parameters (blue and orange) is almost as good as the best performance obtained by doing an evolutionary optimization over the parameters. The latter can be done optimizing NMI for each graph configuration separately (brown) or optimizing the area under the curve (green and red). For graphs of 10,000 nodes the default parameters lead to a poor performance. There is no significant difference between the two different metrics (Euclidean and spherical)}
\label{fig:NMNF}
\end{figure}

The results are in Fig.~\ref{fig:NMNF}. In panel (a) we show the curve obtained with the default parameters above and the ones obtained by optimizing the NMI for each individual graph and for the whole range of $\mu$-values. The optimization is carried out with the procedure described in Section~\ref{sec:GA}.
All curves are better than Infomap's. However, when we use the default parameters for the larger LFR graphs (b), the performance degrades considerably and Infomap does much better. On such graphs we could not derive the optimal curves. Due to the high number of parameters, the optimization is very costly, computationally, so we only report the curve with the default parameters. Louvain offers closer performances on both network sizes, though it is clearly superior for $N=10,000$ and low $mu$-values, where it is capable to detect the correct partition, while MNMF fails to do so consistently.

\subsection{Random Walk Embeddings}

In this Section we present tests carried out by using embeddings that rely on random walks performed on the graph. We chose the two most popular techniques of this class, described below.

\subsubsection{DeepWalk}

\begin{figure}
\includegraphics[width=\columnwidth]{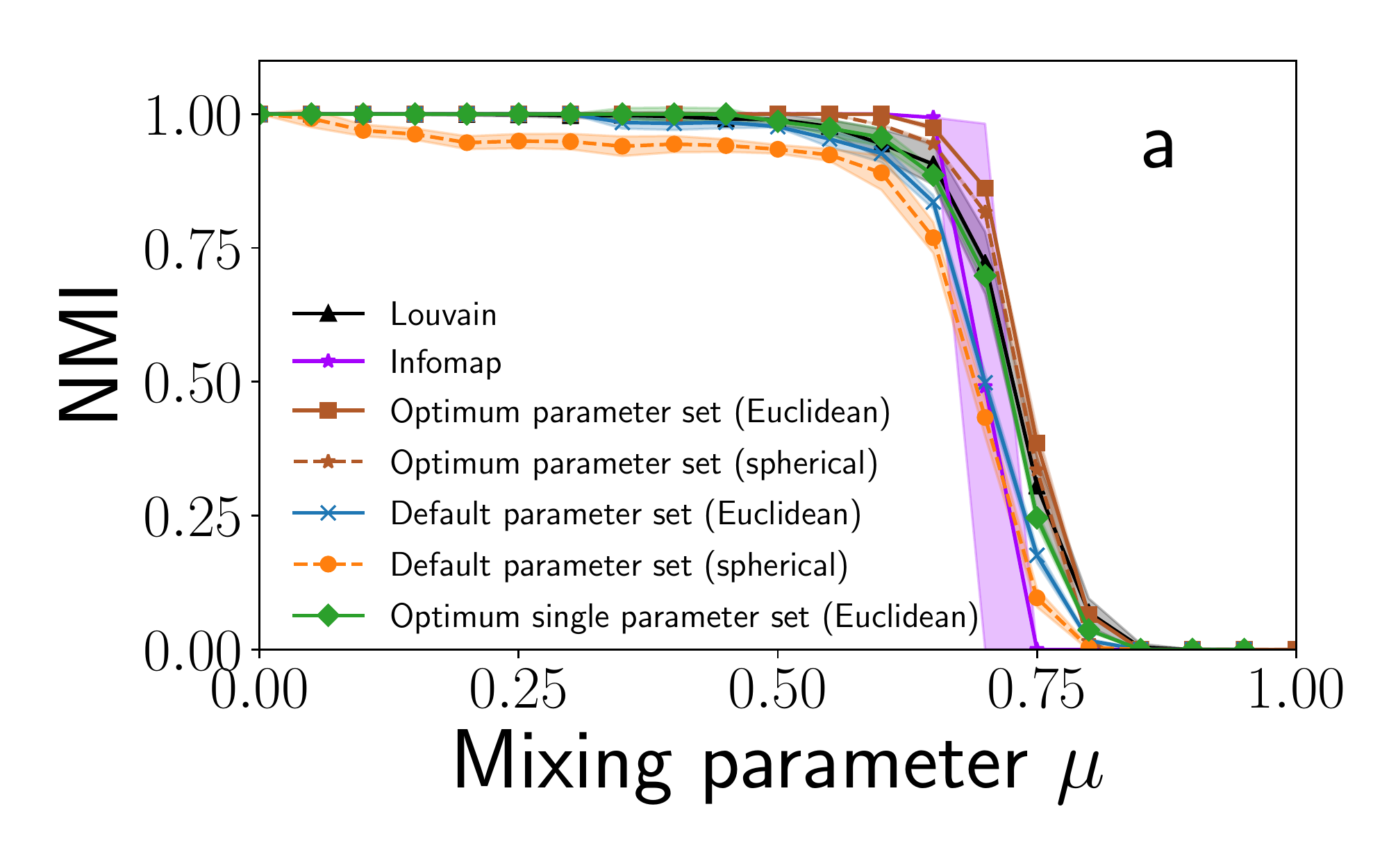}
\includegraphics[width=\columnwidth]{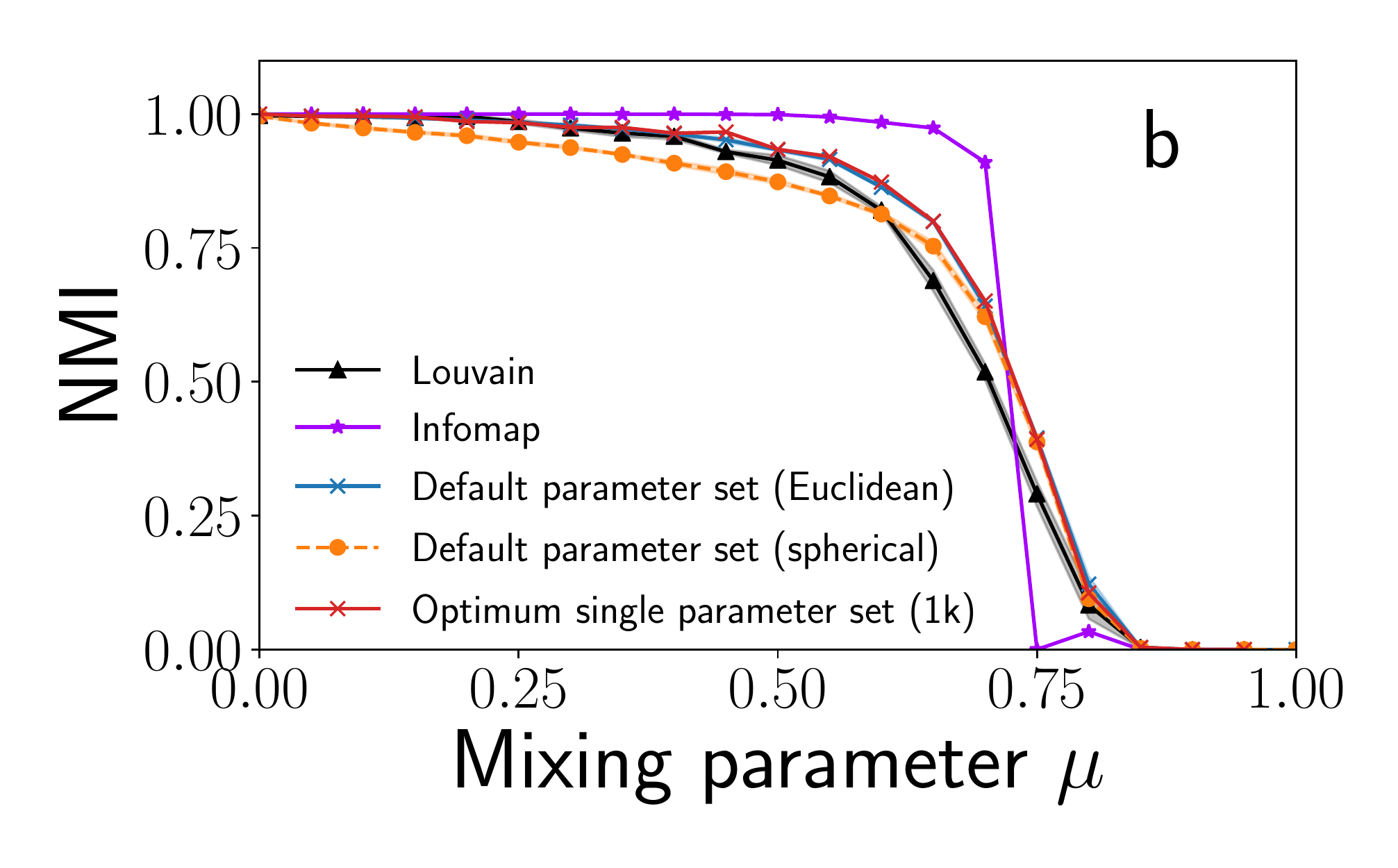}
\caption{Performance of DeepWalk on LFR benchmark graphs of 1,000 nodes (a) and 10,000 nodes (b). The baseline performances are those of Infomap ($\star$) and Louvain (\mytriangle{black}). The performance of the algorithm using the default parameter set is better if we adopt the Euclidean distance (blue, solid) instead of the spherical distance (orange, dashed). The default parameter performance is not significantly worse than the optimum performance found by doing an evolutionary optimization over each configuration (\rule{1.2ex}{1.2ex}) or a single parameter set across all configurations ($\star$). For each value of $\mu$, 10 different LFR graphs are generated and the NMI value shown is the average across them.  }
\label{fig:DW}
\end{figure}

\textit{DeepWalk} \cite{Perozzi2014} applies language modelling techniques from deep learning on graphs instead of words and sentences. The algorithm uses local information obtained from truncated random walks to learn latent representations by treating walks as the equivalent of sentences in the \textit{word2vec} \cite{mikolov13} language model. The default parameters used for the experiments here are: dimension $d = 128$, window $w = 10$, random walk length $t = 40$, number of walks per node $n = 80$.

In Fig.~\ref{fig:DW} we show the performance curves for the default parameters for both standard (Euclidean) and spherical $k$-means, as well as the optimal curves obtained by maximizing the NMI for each single LFR graph and using a single parameter set for the whole range of $\mu$-values. The curves are similar: the default parameter curves are worse than Infomap's and Louvain's, while the optimal performance when the parameters are adjusted for each single network is superior. For the larger LFR graphs (bottom panel) the performance optimization is very expensive, so we only report the curves obtained by using the default parameters and the optimal parameter set found on the smaller graphs. The latter closely follows the curve of the default parameters (Euclidean distance). Infomap is clearly superior here, while Louvain has comparable performance as DeepWalk's default curves.

\begin{figure}
\includegraphics[width=\columnwidth]{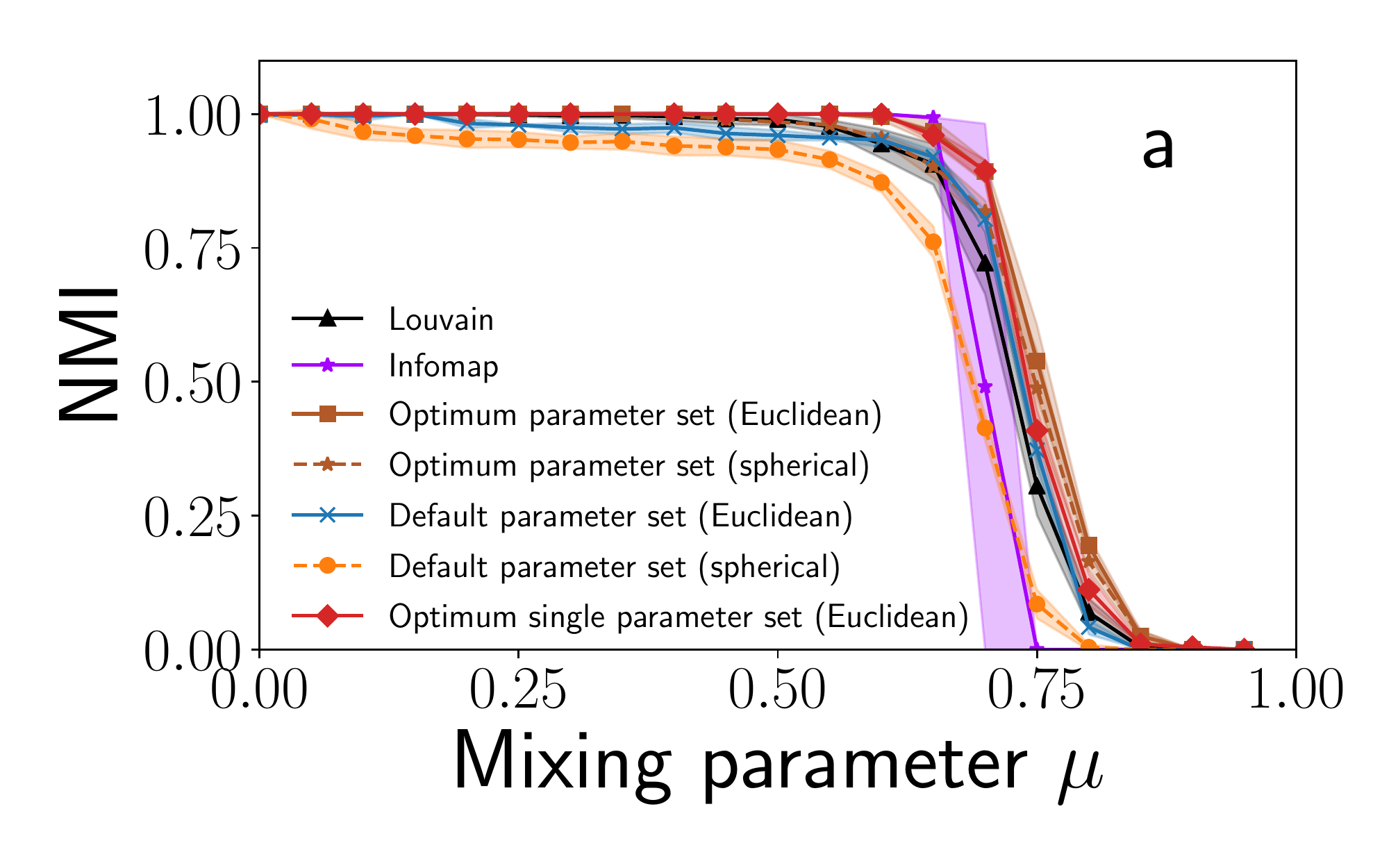}
\includegraphics[width=\columnwidth]{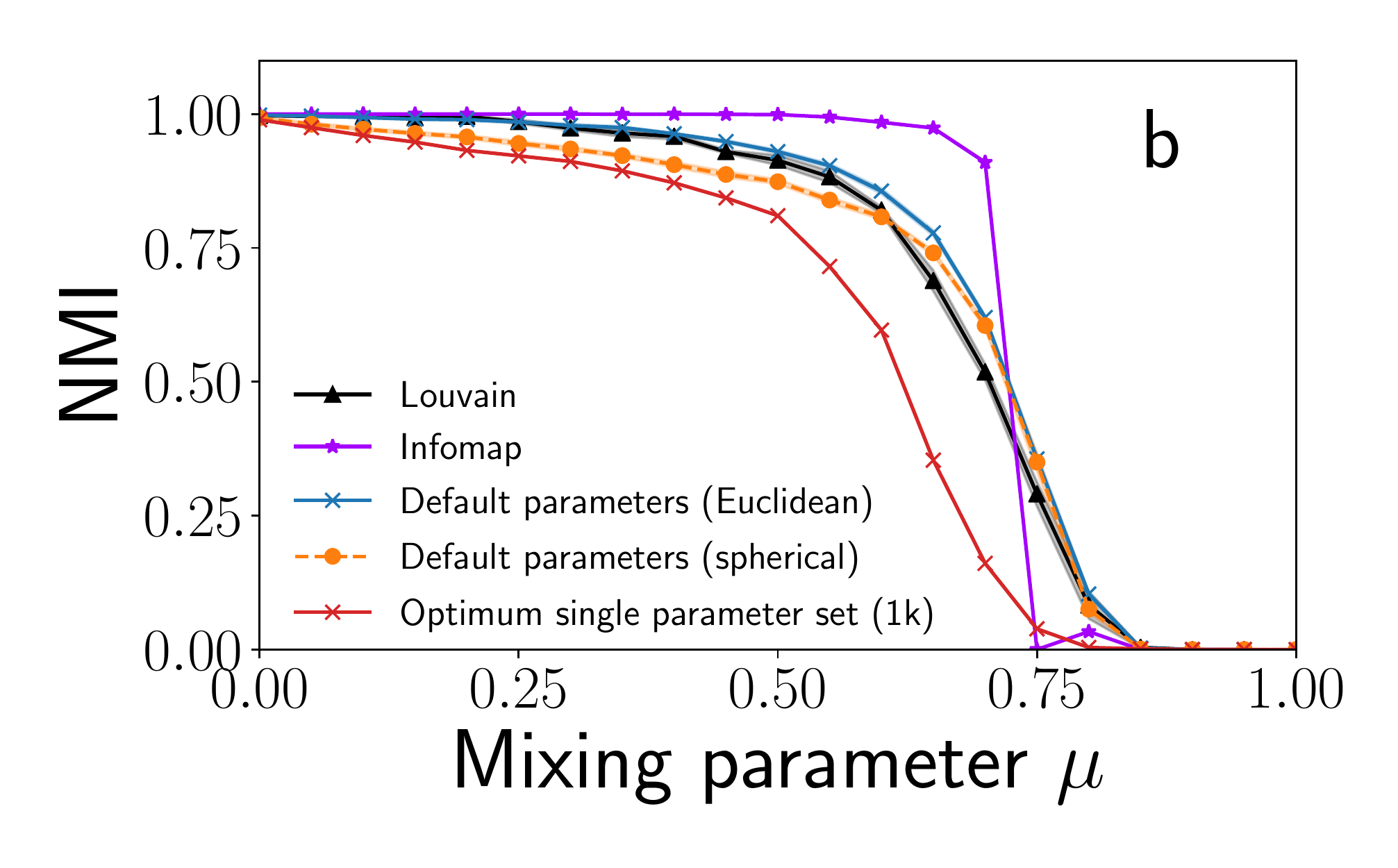}
\caption{Performance of node2vec on LFR benchmark graphs of 1,000 nodes (a) and 10,000 nodes (b). The baseline performances are those of Infomap ($\star$) and Louvain (\mytriangle{black}). The performance of the algorithm using the default parameter set is better if we adopt the Euclidean distance (blue, solid) instead of the spherical distance (orange, dashed). The default parameter performance is not significantly worse than the optimum performance found by doing an evolutionary optimization over each configuration (\rule{1.2ex}{1.2ex}) or a single parameter set across all configurations ($\star$). For each value of $\mu$, 10 different LFR graphs are generated and the NMI value shown is the average across them. }
\label{fig:N2V}
\end{figure}

\subsubsection{node2vec}
 
\textit{node2vec}~\cite{Grover2016} uses the same optimization procedure as DeepWalk, but the process to generate the "sentences" is different. Instead of using simple random walks as in DeepWalk, node2vec uses biased random walks. The walk consists of a mixture of steps following breadth-first and depth-first search, with parameters $p$ and $q$ regulating the relative weights of the two approaches. The default parameters used for the experiments here are: dimension $d = 128$, window $w = 10$, random walk length $t = 10$, number of walks per node $n = 80$, biased walk weights $p = 1$ and $q = 1$.

In Fig.~\ref{fig:N2V} we show the performance curves for the default parameters for both standard (Euclidean) and spherical $k$-means, as well as the optimal curves obtained by maximizing the NMI for each single LFR graph and using a single parameter set for the whole range of $\mu$-values. For the default parameters the results are very similar as for DeepWalk, since for those parameters the two techniques are equivalent (Fig.~\ref{fig:DW}). 
For the larger LFR graphs we include the curve with the optimal single parameter set found for the smaller networks, which is significantly worse than the default parameter curves in this case. This shows that the optimal parameter set is strongly dependent of the network size, along with other features.

\subsection{Large-Scale Information Network Embedding (LINE)}

\textit{Large-Scale Information Network Embedding} (LINE)~\cite{Tang2015} projects nodes the closer to each other the higher their similarity. It considers both first order similarity, based on whether nodes are adjacent or not, and second order similarity, based on the overlap of the neighborhoods of two nodes. 

There are two different ways to obtain the embeddings, denoted LINE-1 and LINE-2. The authors also suggest concatenating the two embeddings to obtain a $2d$ dimensional embedding which we denote by LINE-1+2. 

\begin{figure}
\includegraphics[width=\columnwidth]{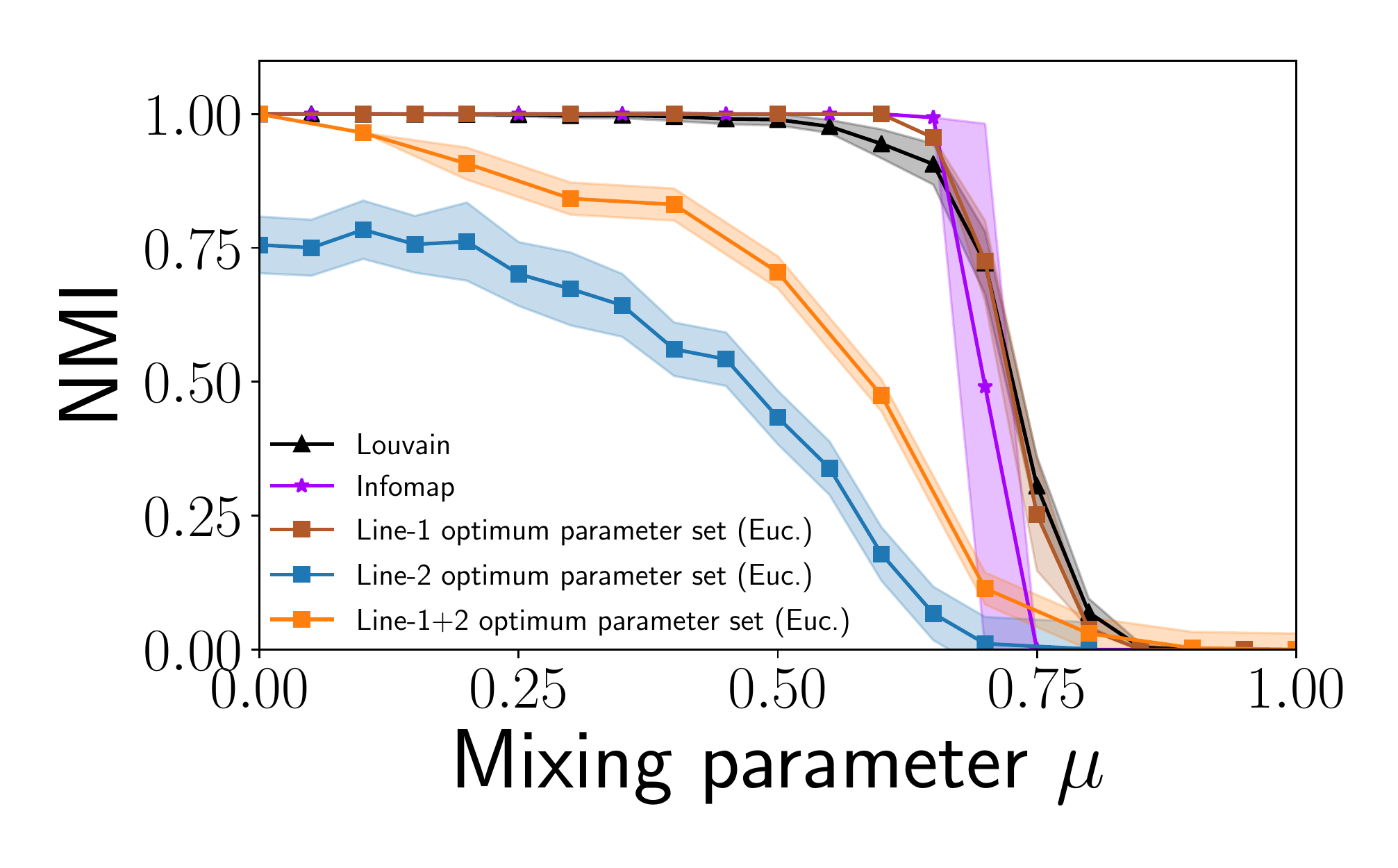}
\caption{Performance of Large-Scale Information Network Embedding on LFR benchmark graphs for 1,000 nodes. The baseline performances are those ofInfomap ($\star$) and Louvain (\mytriangle{black}). The curves show the optimal performance of the three implementations of the method. The first implementation, LINE-1, is clearly the best and outperforms Infomap, albeit by a small margin. }
\label{fig:LINE}
\end{figure}

In Fig.~\ref{fig:LINE} we show the results for the LFR graphs with 1,000 nodes. The method is very slow so we could not produce the analogous curve for benchmark graphs with 10,000 nodes. For each of the three different implementations of the method we have derived the optimum parameters via the genetic algorithm of Section~\ref{sec:GA}, for each LFR network. For LINE-1 the optimal performance exceeds those of Infomap and Louvain, though the curves are very close, the other two implementations are far worse. We do not show the curves corresponding to the default parameters because they are very poor.

In Tables I and II we summarize our comparative analysis by reporting the performances of the various techniques, measured via the area under the curve.

\begin{table*}
\centering
\begin{tabular}{ |p{3cm}||p{2cm}|p{2cm}|p{2cm}|  }
 \hline
 \multicolumn{4}{|c|}{Area Under Curve} \\
 \hline
Algorithm & Default parameter set & Optimum parameter set & Optimum single parameter set\\
\hline
 Infomap   & 0.60   & & \\
 \cline{1-4}
  Louvain   & \textbf{0.62}   & & \\
  \hline
 Laplacian Eigenmap &   & 0.64     &\\
\hline
 LLE &  & 0.64  & \\
\hline
 HOPE &  & 0.64 & 0.62 \\
\hline
 M-NMF    & 0.63 & 0.65 &  0.64  \\
\hline
 DeepWalk & 0.59 & 0.64  & 0.61   \\
\hline
 node2vec & 0.62  &  0.66 &  0.64  \\
\hline
 LINE &  &  0.62    &\\
 \hline
\end{tabular}
\caption{Comparison of the performances of Infomap and the embedding-based clustering algorithms we have adopted in our tests, for LFR benchmark graphs of $1,000$ nodes. Performance is estimated by computing the area under the curve (AUC). We started computing the AUC from $\mu=0.1$ because in some cases we do not have results for lower $\mu$-values. Each value in the table is the higher score between the one obtained using the Euclidean distance and the one obtained using the spherical distance. The AUC of the best traditional community detection method is indicated in boldface.}
\end{table*}

\begin{table*}
\centering
\begin{tabular}{ |p{3cm}||p{2cm}|p{2cm}| }
 \hline
 \multicolumn{3}{|c|}{Area Under Curve} \\
 \hline
Algorithm & Default parameter set & Optimum parameter set \\
\hline
 Infomap   & \textbf{0.62}  & \\
 \cline{1-3}
  Louvain   & 0.58  & \\
  \hline
 Laplacian Eigenmap &   & 0.66    \\
\hline
 LLE &  & 0.64   \\
\hline
 M-NMF    & 0.57  &   \\
\hline
 DeepWalk & 0.60 &     \\
\hline
 node2vec & 0.59  &     \\
 \hline
\end{tabular}
\caption{Same as Table I, but for LFR benchmark graphs of $10,000$ nodes. The AUC of the best traditional community detection method is indicated in boldface.}
\end{table*}

\section{Conclusions}

We have evaluated the perfomance of graph clustering techniques mediated by embeddings of networks in high-dimensional vector spaces. The identification of the clusters in the vector space is done via data clustering methods, specifically $k$-means.
Overall we found that the several embedding strategies we have adopted do not help to resolve the community structure of LFR benchmark graphs better than the best performing community detection algorithms, Infomap and Louvain, which act directly on the network, without any embedding. The parameters of the embedding procedure can be optimized such to get close or even outperform the curve of those algorithms. However, the optimal parameters generally vary with the mixing parameter $\mu$, so we could not come up with a single parameter set that we can recommend for clustering applications. Besides, the optimal parameter values are affected by the network size as well, so there would not be a "one-size-fits-all" parameter set. This means that, for a given real network, we cannot know which parameters are best to reveal its modular structure, which results in noisy partitions.  
Finally, the combination of embedding techniques plus data clustering is a computationally expensive procedure. While some embedding algorithms can scale up to very large graphs, data clustering techniques (like $k$-means) typically scale superlinearly with the graph size.  As a result, the full procedure is much more computationally demanding than fast graph clustering methods, like Infomap and Louvain.

We acknowledge that the embedding is just a component of the overall clustering algorithm, and that the performances we observe might be due to the data clustering approach used to group the points into clusters. We have used $k$-means and Gaussian Mixture Models (results not shown), which are regularly adopted for data clustering, without finding significant differences. Still, the number of clusters, which is usually unknown in practice,
needs to be specified as input, and we set it equal to the correct value for the planted partition. By doing that we have significantly helped the performance, whereas many network clustering algorithms (including Infomap and Louvain) are able to infer the number of clusters. This means that the curves we have shown in our plots are better, in general, than the ones we would obtain if the number of clusters were inferred via some criterion.
We also recognize that the concept of distance becomes problematic in high-dimensional spaces and work is in progress to alleviate the drawbacks deriving from that.

Embedding techniques have not been designed to tackle specifically the clustering task, so it is not surprising that they do not excel in this task. To improve their clustering performance embedding strategies that focus on preserving the modular structure of networks should be developed. Spectral-based embeddings seem particularly promising in that regard, because of the provable optimal performance of spectral clustering in synthetic graphs with communities built with stochastic blockmodels~\cite{peixoto20} and because they provide ways to estimate the number of clusters~\cite{qin13,zhao12b,saade14,dallamico19,krzakala13}.

\section{Acknowledgments}
This project was funded by the Deanship of Scientific Research (DSR) at King Abdulaziz University, Jeddah, Saudi Arabia, under Grant No. RG-1439-311-10. AA, VT, WA and SF therefore, acknowledge with thanks DSR for technical and financial support. FR acknowledges partial support from the National Science Foundation (CMMI-1552487).

\bibliography{embedding}
\clearpage
\appendix

\section{Comparative analysis of embedding methods on SBM's}

\begin{figure*}[h]
    \centering
    \includegraphics[width = 0.85\textwidth]{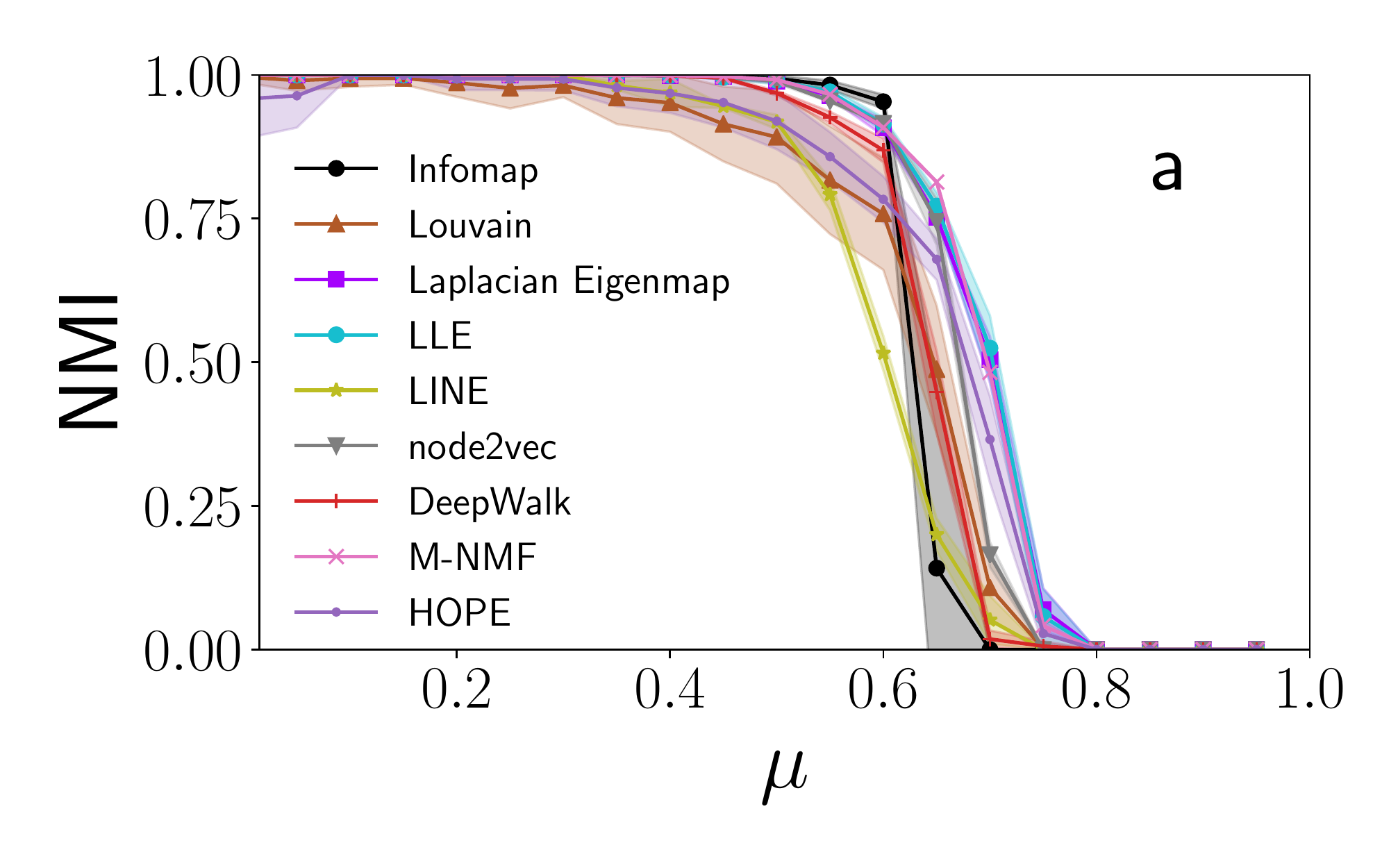}
    \includegraphics[width = 0.85\textwidth]{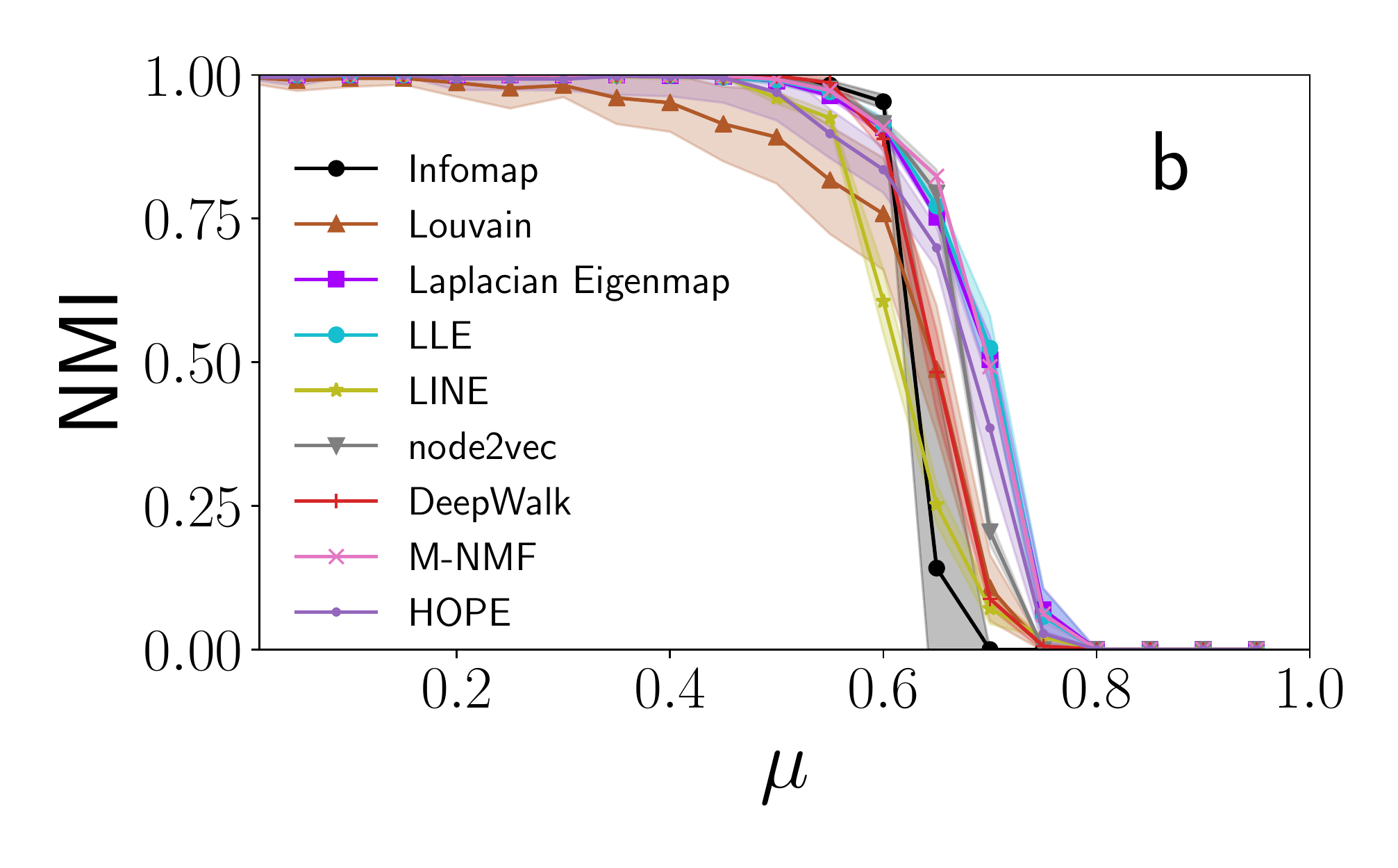}
    \caption{Performance of all the considered embedding methods on a stochastic block model (SBM) of 20 groups of 50 nodes each. The average degree is set to 20. The mixing parameter $\mu$ is reported on the $x$-axis, for each $\mu$-value ten graphs are generated and the average NMI over them is shown on the y-axis. Panel (a) shows the curves corresponding to the default parameters, panel (b) the best ones obtained via the genetic optimization over the parameters (Sec.~\ref{sec:GA}).}
    \label{fig:SBM20x50}
\end{figure*}

\begin{figure*}[h]
    \centering
    \includegraphics[width = 0.85\textwidth]{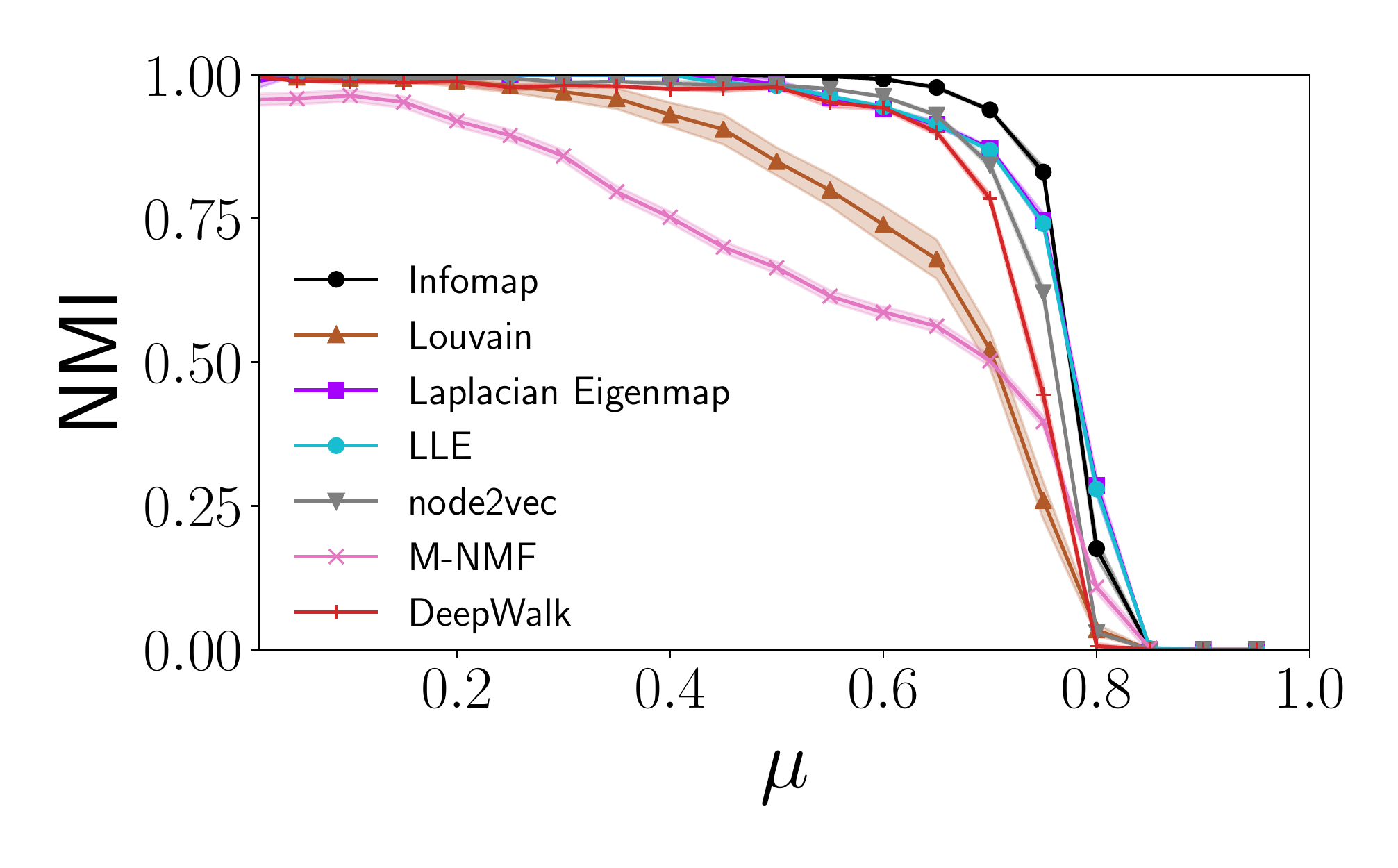}
    \caption{Performance of all the considered embedding methods on a stochastic block model (SBM) of 200 groups of 50 nodes each. The average degree is set to 20. The mixing parameter $\mu$ is reported on the $x$-axis, for each $\mu$-value ten graphs are generated and the average NMI over them is shown on the y-axis for the default parameters.}
    \label{fig:SBM200x50}
\end{figure*}

We also performed tests on networks generated by the classic \textit{stochastic blockmodel} (SBM)~\cite{peixoto20}. The SBM is a model of graphs with built-in community structure. The probability to form a link between two nodes only depends on the groups the nodes belong to. The LFR benchmark is a special case of SBM. We considered graphs with 20 groups with 50 nodes each, the average degree is 20. The mixing parameter $\mu$ is again the ratio between the external degree of a node and its total degree. The best parameters of the embedding methods are chosen after genetic optimization in the manner outlined in Sec.~\ref{sec:GA}: the optimum NMI curves are plotted in Fig.~\ref{fig:SBM20x50}b. Like for the LFR benchmark tests, the performance curves of Infomap and Louvain are shown for comparison.  Embedding techniques do a bit better there (except LINE). In Fig.~\ref{fig:SBM200x50} we show the performance curves of the embedding clustering methods for graphs with $10,000$ nodes, with $200$ groups of $50$ nodes each. Average degree is still $20$. For LE and LLE the number of dimensions, their only parameter, is set to the number of clusters, $200$. We see that the embedding methods' performance worsens a bit here, while Infomap improves and outperforms all of them. Louvain's performance degrades considerably.
We remind that the correct number of clusters is fed into the embedding-based methods, in contrast to the standard clustering techniques, which are capable to guess it. This confers a major advantage to embedding-based clustering methods over the traditional ones.

%

\section{Comparison of traditional community detection methods}

Here we compare the performance of four clustering algorithms on the LFR benchmark. Besides Infomap and Louvain the other algorithms are:
\begin{itemize}
    \item The Order Statistics Local Optimization Method (OSLOM)~\cite{lancichinetti11} seeks statistically significant clusters in networks. We use the OSLOM code available at \href{http://oslom.org/}{\textit{oslom.org}}, with default parameters. 
    \item The Label Propagation Algorithm (LPA)~\cite{raghavan07}. It is a technique that assigns nodes to the community to which the majority of its neighbors belongs to. The code was taken from the \href{http://igraph.org/}{\textit{igraph}} library.
\end{itemize}
 
\begin{figure*}
    \centering
    \includegraphics[width = 0.85\textwidth]{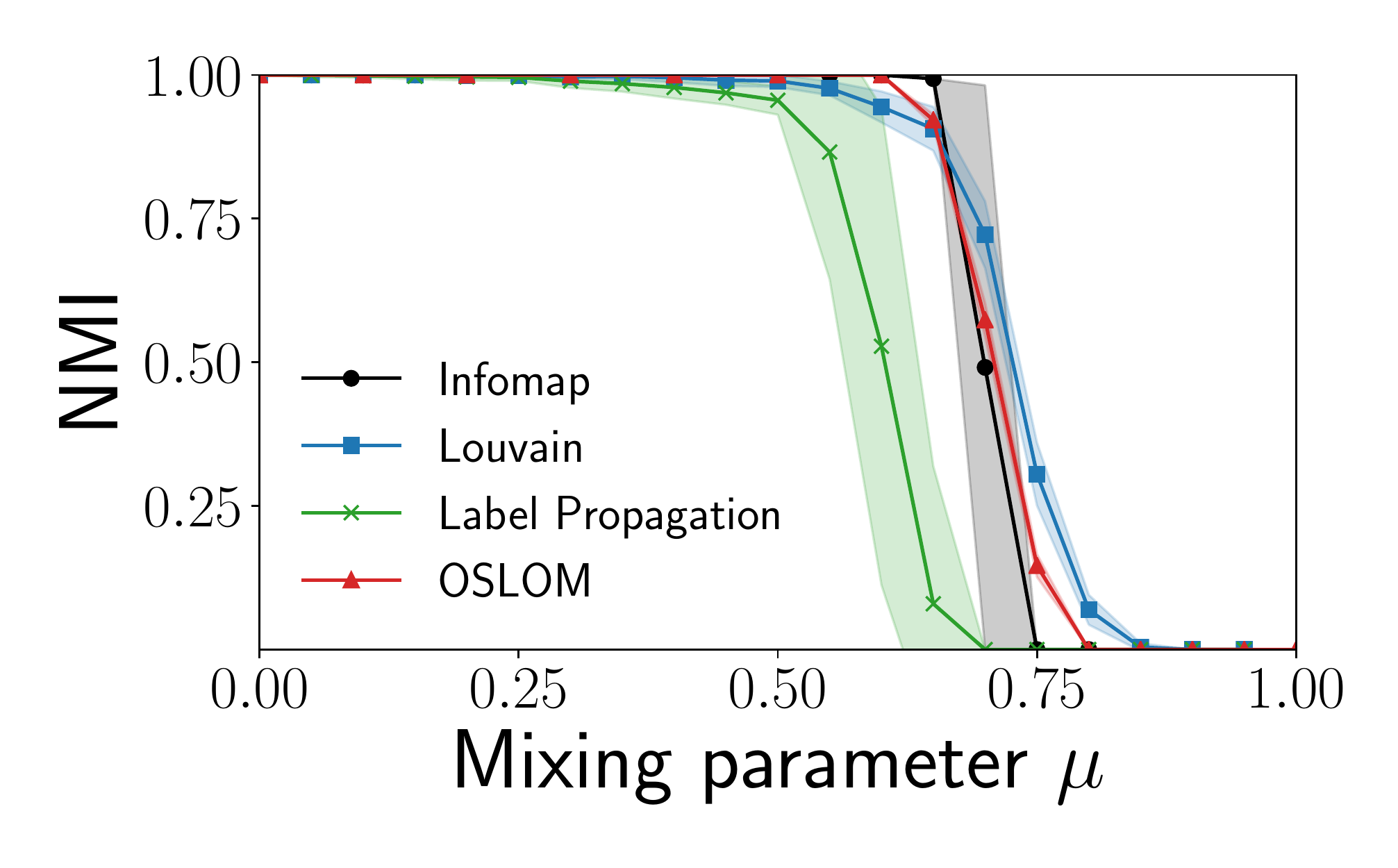}
    \caption{A comparison of traditional community detection methods --- Infomap, Louvain, Label Propagation and OSLOM on the LFR benchmark. The average degree, $k = 20$, and number of nodes $N = 1000$. The additional parameters for the LFR benchmark are $\tau_1 = 2$,
    $\tau_2 = 3$,
    $k_{max} = 50$,
    $c_{min} = 10$, and
    $c_{max} = 100$. }
    \label{fig:CDM}
\end{figure*}

Fig.~\ref{fig:CDM} shows that Infomap, Louvain and OSLOM have comparable performance, whereas label propagation is worse. We used Infomap and Louvain in the tests of the main text as representative of well-performing algorithms on the LFR benchmarks, and for no other reason.

\section{Failure of spectral methods at high dimensions for $\mu \rightarrow 0$}

\begin{figure*}
    \centering
    \includegraphics[width=0.85\textwidth]{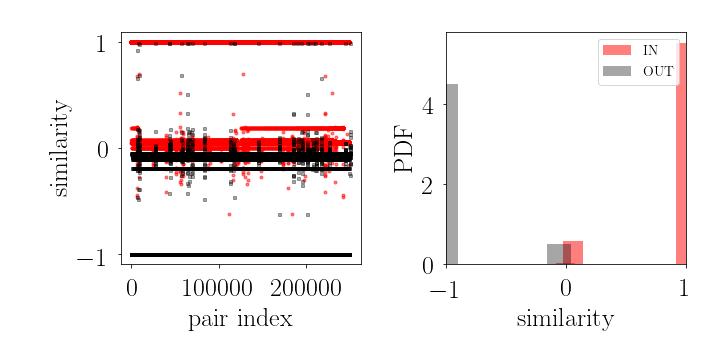}
    \includegraphics[width=0.85\textwidth]{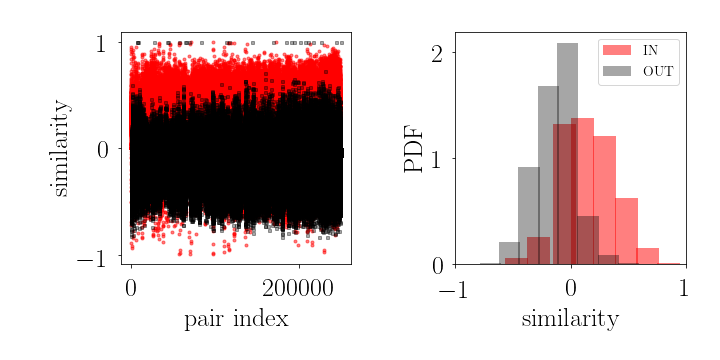}
    \includegraphics[width=0.85\textwidth]{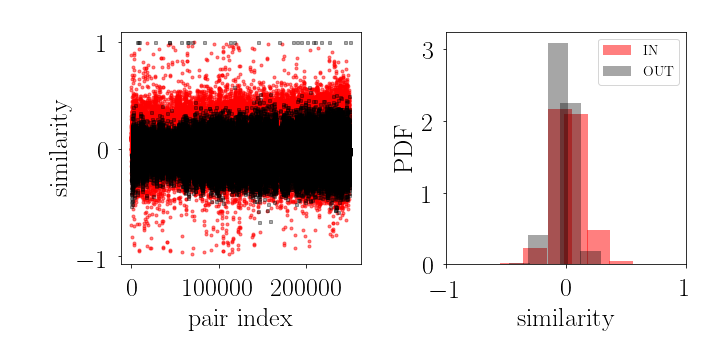}
    \caption{Problem of spectral embeddings for well-separated communities. The network is generated by an SBM with two blocks, $p_{\textrm{in}} = 1$ and  $p_{\textrm{out}} = 0.0001$ and N=500 in each block. Laplacian Eigenmap is applied for dimensions $d = 20,50,100$ (top to bottom). The cosine similarity for within cluster node pairs overlaps with that of the node pairs across clusters for higher embedding dimension, making the identification of the clusters hard even for this straightforward example.}
    \label{fig:LE1}
\end{figure*}

\begin{figure*}
    \centering
    \includegraphics[width=0.85\textwidth]{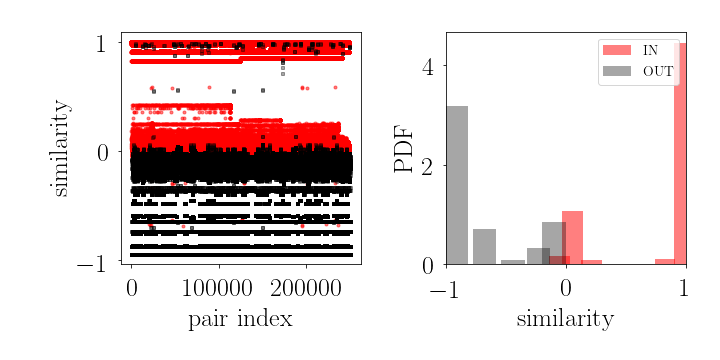}
    \includegraphics[width=0.85\textwidth]{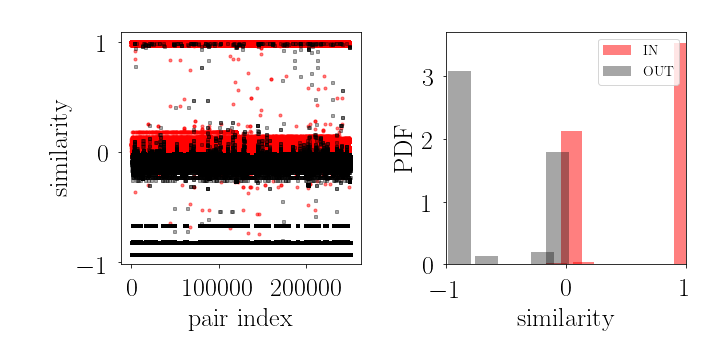}
    \includegraphics[width=0.85\textwidth]{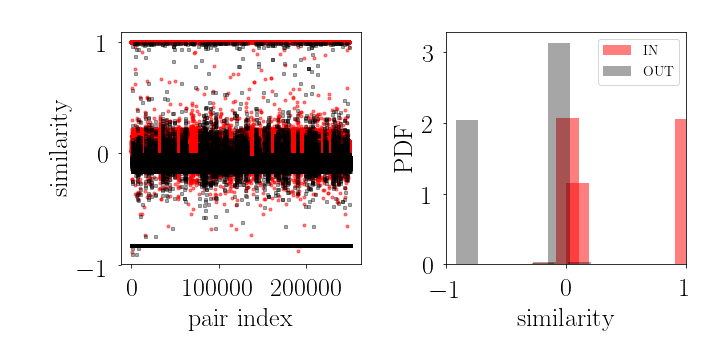}
    \caption{Problem of spectral embeddings for well-separated communities. The network is generated by an SBM with two blocks, $p_{\textrm{in}} = 1$ and  $p_{\textrm{out}} = 0.001$ and N=500 in each block. Laplacian Eigenmap is applied for dimensions $d = 20,50,100$ (top to bottom). The cosine similarity for within cluster node pairs has limited overlap with that of the node pairs across clusters for higher embedding dimension, making the identification of the clusters easier.}
    \label{fig:LE1a}
\end{figure*}

\begin{figure*}
    \centering
    \includegraphics[width=0.85\textwidth]{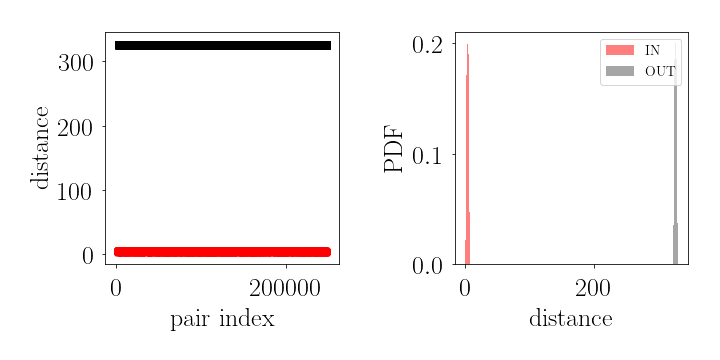}
    \includegraphics[width =0.85\textwidth]{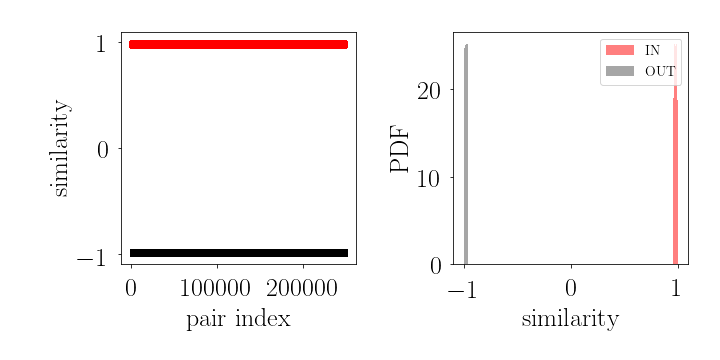}
    \caption{Improving Laplacian Eigenmap. The network is generated by the same SBM as in Fig~\ref{fig:LE1}. A modified version of Laplacian Eigenmap for $d=100$ is applied, in that the eigenvector components are multiplied by the (inverse) eigenvalue. In the first row we show analogous diagrams as in Figs~\ref{fig:LE1} and \ref{fig:LE1a}, for the Euclidean distance (top row) and the cosine similarity (bottom row). The clusters are now well separated.} 
    \label{fig:LE3}
\end{figure*}

In Figs.~\ref{fig:LE} and \ref{fig:LLE} we have seen that the performance for spectral embedding methods like Laplacian Eigenmap and Locally Linear Embedding is not very good for low values of $\mu$. This is against intuition, as clusters there are well separated from each other and fairly easy to resolve. To check what happens we have considered a relatively simple graph with two communities, built via the \textit{stochastic blockmodel} (SBM)~\cite{peixoto20}. Here we consider only two communities, with 500 nodes each. The probability of having links within the groups is $p_{in}=1$, the probability of having links between the groups is $p_{out}=0.0001$, so that the communities are well separated from each other. In this example we expect that the embedding generates two sets of points well separated from each other. In Fig.~\ref{fig:LE1} we see what happens if we do the embedding of this network using Laplacian Eigenmap for different numbers of dimensions: $d=20$, $50$, $100$. Each row of plots refers to a value of $d$ (growing from top to bottom). For each $d$, we compute the cosine similarity of all pairs of nodes, computed via their respective vectors. Red and black/grey indicate pairs in the same versus different groups. The expectation is that node pairs in the same cluster have higher similarity than pairs in different clusters. The left plot shows the actual similarity values for each pair and indeed we see that pairs of nodes in the same community are very similar, while pairs of nodes in different communities are very dissimilar. This is further indicated by the right diagram, showing the probability distribution of the similarities. We see that, as $d$ increases, the difference between within-community and between-community pairs reduces. For $d=100$ pairs of nodes have very low similarity, regardless of their group memberships, which makes it difficult to separate the nodes. This is due to the peculiar behavior of the eigenvectors of the Laplacian when the clusters are almost disjoint. On the other hand, in Fig.~\ref{fig:LE1a}, we consider an SBM like that in Fig.~\ref{fig:LE1}, but with $p_{in}=1$ and $p_{out}=0.001$. Now we see that the two groups can be identified even at high dimensions. In the special case of Laplacian Eigenmap, we can modify the embedding strategy by multiplying the eigenvector components by the inverse of the corresponding eigenvalue. By doing that we see that the two groups are clearly separated (Fig.~\ref{fig:LE3}). This trick, however, cannot be easily extended to other embedding strategies relying on graph spectra.


\end{document}